\documentclass[aps,prl,floats,reprint,superscriptaddress]{revtex4-1}
\usepackage{amssymb}
\usepackage{amsmath}
\usepackage{graphicx}

 \def\bc{\begin{center}}          \def\ec{\end{center}}
 \allowdisplaybreaks
\begin{document}
 \title{Guiding femtosecond high-intensity high-contrast laser pulses by copper capillaries}
 \author{V.E.Leshchenko}
 \affiliation{Institute of Laser Physics SB RAS, 630090, Novosibirsk, Russia}
 \affiliation{Novosibirsk State University, 630090, Novosibirsk, Russia}
 \author{K.V.Gubin}
 \affiliation{Institute of Laser Physics SB RAS, 630090, Novosibirsk, Russia}
 \author{V.I.Trunov}
 \affiliation{Institute of Laser Physics SB RAS, 630090, Novosibirsk, Russia}
 \affiliation{Novosibirsk State University, 630090, Novosibirsk, Russia}
 \author{K.V.Lotov}
 \affiliation{Novosibirsk State University, 630090, Novosibirsk, Russia}
 \affiliation{Budker Institute of Nuclear Physics SB RAS, 630090, Novosibirsk, Russia}
 \author{E.V.Pestryakov}
 \affiliation{Institute of Laser Physics SB RAS, 630090, Novosibirsk, Russia}
 \affiliation{Novosibirsk State University, 630090, Novosibirsk, Russia}
 \date{\today}
 \begin{abstract}
Propagation of high-intensity, high-contrast ($<10^{-8}$), 50\,fs laser pulses through triangilar copper capillaries is experimentally studied. The relative transmission through 20-mm-long, about 50\,$\mu$m wide capillaries is directly measured to be 70\% for input intensities up to $10^{17}\,\text{W/cm}^2$. The copper reflectivity in vacuum, helium, and air is measured in the intensity range of $10^{10}$--$10^{17}\,\text{W/cm}^2$. No reflectivity decrease in vacuum and helium is observed, which leads to the conclusion that copper capillary waveguides can efficiently guide laser pulses of intensities up to $10^{19}\,\text{W/cm}^2$ on the capillary axis (that corresponds to  $10^{17}\,\text{W/cm}^2$ on the walls). The reduction of the transmission efficiency to zero after a number of transmitted pulses is observed, which is caused by plug formation inside the capillary. The dependence of the capillary lifetime on the pulse energy is measured.
 \end{abstract}
 \maketitle

The laser wakefield acceleration in a plasma \cite{RMP81-1229} is now recognized as a perspective method of electron acceleration to GeV range energies in table-top devices. These electrons can be mainly used in advanced light sources \cite{RMP85-1,PPCF56-084015}; the high energy physics is also considered as a future area of their application \cite{PRST-AB15-051301}.

The intense laser pulses must propagate many Rayleigh lengths in the plasma to accelerate electrons to high energies. Since the relativistic self-focusing is ineffective in preventing the diffraction of short laser pulses \cite{RMP81-1229}, various guiding methods are proposed to increase the propagation length. The most widely used one is based on preformed plasma channels with the density minimum on the axis \cite{RMP81-1229}. The record energy of accelerated electrons (4.2\,GeV) was demonstrated using this method \cite{PRL113-245002}.

An alternative approach is based on the use of capillary tubes with the inner radius of the order of the laser pulse waist. The capillary walls guide the laser pulse, while the plasma inside the capillary supports the plasma wave \cite{PAc63-139,LPB19-219}. The plasma is created by optical field ionization of the gas that fills the capillary. Note that a similar setup can be used for direct X-ray generation without the electron beam as an intermediate agent \cite{LPB9-725,PRA73-033801}.

Both dielectric \cite{OptLett20-1086,PRL82-4655,PRL83-2187,JOSAB27-1400,PoP19-093121, APB105-309,PoP20-083106,PRST-AB17-051302} and plasma-walled \cite{OptLett20-1086,PRE57-4899,PRL92-205002} capillaries have been studied in this context. Dielectric capillaries typically have the diameter larger than 100\,$\mu$m, and the pulse intensity for them is limited to avoid ionization of the walls. In contrast, plasma-walled capillaries need a contact of a high intensity laser pulse with the walls and are tens of microns in diameter. Metal capillaries can be considered as a special case of plasma-walled capillaries, as the metal walls behave like a cold dense plasma with a sharp boundary. Laser propagation in dielectric capillaries is understood quite well \cite{PoP20-083120,PRE86-066411}, and experimental observations agree with the theory \cite{JOSAB27-1400,PoP8-3445}. Plasma-walled capillaries are more difficult for theoretical analysis, the available theory \cite{PRE64-016404,PoP13-053114,TP49-91} is not experimentally verified yet, the experiments were performed in the sub-picosecond regime \cite{OptLett20-1086,PRE57-4899,PRL92-205002}, and no experimental data are available for femtosecond pulses. Nevertheless, the plasma-walled capillaries are of a certain interest as they have the potential to squeeze the laser pulse most tightly, as compared to other channeling schemes, and to control its trajectory precisely. Since the pulse intensity on the walls is no more limited by the ablation risk, there is also no strict limitation on the on-axis intensity. If combined with techniques of fast capillary fabrication and replacement, this scheme may open a path to minimization of the driver power required for production of electrons of a designated energy \cite{LPB19-219}. In this paper, we present the first experimental study of short laser pulse propagation through ablating copper capillaries and capillary degradation due to the ablation; both effects are crucial for realizability of the laser wakefield acceleration in plasma-walled capillaries.

\begin{figure}[b]
\includegraphics[width=80mm]{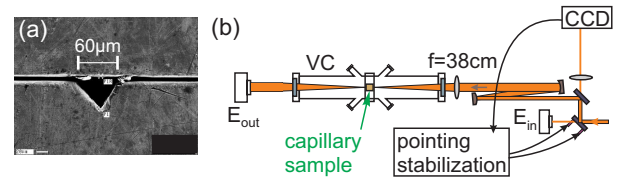}
\caption{ (a) SEM image of the capillary inlet. (b) Experimental setup (VC is the vacuum chamber).}\label{fig1}
\end{figure}
The experiments are performed using the laser system created at the Institute of Laser Physics SB RAS \cite{LPL11-095301}. The system is based on the parametric amplification under picosecond ($\sim$90\,psec) pumping. It delivers up to 30\,mJ, 50\,fs, pointing stabilized pulses at 850 nm central wavelength with a high time contrast ($10^{-8}$) to the capillary entrance. The capillaries are produced by engraving triangular grooves on a polished copper plate; another (not grooved) plate closes the capillary [Fig.\,\ref{fig1}(a)]. The advantage of the triangular shape is its manufacturability, disadvantages are the complexity of theoretical analysis and an unknown mode structure. The capillary quality is controlled with a scanning electron microscope (SEM). The capillary length is 20\,mm, the side of the triangle is about 60 $\mu$m. The apparent black line between the copper plates in Fig.\,\ref{fig1}(a), that could be misinterpreted as a gap, is caused by small cants on the edges of the plates. The real gap between the plates is measured to be smaller than 0.5\,$\mu$m. The experimental setup is shown in Fig.\,1(b). The output energy ($E_\text{out}$) is registered by the pyroelectric energy meter (PE50BF, Ophir). To determine the input energy ($E_\text{in}$), a small part of the pulse leaked through the mirror is registered by another pyroelectric energy meter (PE10-V2, Ophir). Samples with capillaries are mounted in the vacuum chamber (VC) with the pressure of $10^{-5}$ Bar. The vacuum chamber is located on the five-dimensional stage (3 coordinates and 2 angles) and rest on the same optical table as the laser to minimize the beam instability.

\begin{figure}[tb]
\includegraphics[width=80mm]{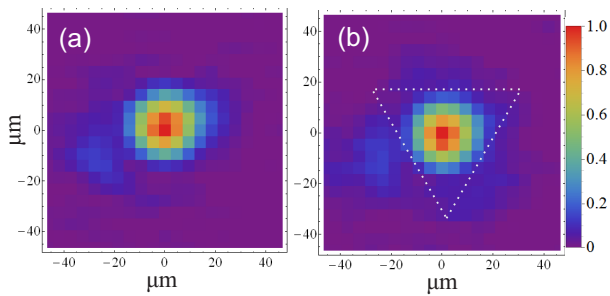}
\caption{ Far field distributions for 1\,mJ (a) and 30\,mJ (b) input energies. The  dotted line shows the input aperture used to calculate the relative efficiency.}\label{fig2}
\end{figure}

In this setup, the focused pulses propagate through the material (fused silica) of the lens and the input window which causes some nonlinear distortions. To minimize these distortions, thin optical elements are used: the lens is 4\,mm thick, and the window is 3\,mm thick. Moreover, the beam is about two times expanded by the reflective telescope based on spherical mirrors to decrease the intensity by increasing the beam size to the value limited by the window aperture. As a result, there are almost no focal spot changes as the energy increases (Fig.\,\ref{fig2}), and the spectrum demonstrates no presence of self-phase modulation. The pulse profile at the far field (Fig.\,\ref{fig2}) has the nearly Gaussian shape without hot spots, as monitored with the WinCamD camera with $4.6 \times 4.6\,\mu$m resolution. The focal spot size is experimentally adjusted for the maximum transmission through the capillary by optimizing the focal length of the lens. The best results are achieved with the focal spot of 35\,$\mu$m in diameter (on the $e^{-2}$ level) using the lens with the focal length of 38\,cm.

\begin{figure}[tb]
\includegraphics[width=80mm]{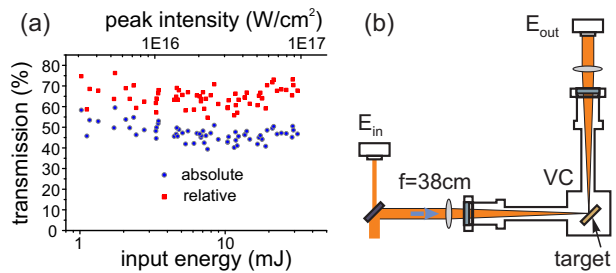}
\caption{ (a) The dependence of capillary transmission on the pulse energy or peak intensity. (b) The experimental setup for measurements of the copper reflectivity. }\label{fig3}
\end{figure}

The measured dependence of the capillary transmission on the pulse energy is shown in Fig.\,\ref{fig3}(a). The absolute transmission is the ratio between output and incident energies, the relative one is the ratio between the output energy and the fraction of the incident energy that falls into the capillary [that is, the energy inside the dotted triangle in Fig.\,\ref{fig2}(b)]. Thus, the relative transmission characterizes the capillary transmission itself excluding the energy losses at the entrance. The data show no relative transmission changes. The absolute transmission slightly decreases from 55-60\% at 1\,mJ input energy to 50\% at 30\,mJ, which is caused by a slight far field profile degradation because of nonlinear distortions. The relative transmission at the maximum available incident energy (30\,mJ) almost coincides with that for unamplified pulses (0.5\,nJ). The experiments show no signs of blocking the capillary entrance by the plasma appearing due to the prepulse action, which was earlier considered as a serious drawback of plasma-walled capillaries.

At the maximum pulse energy of 30\,mJ, experiments with helium-filled capilaries are also performed, and no transmission change is observed as the helium pressure increases from $10^{-5}$ to 1\,Bar. This demonstrates the potential ability to fill the capillary with a gas that has the pressure required for the optimal plasma wave excitation and does not affect the capillary transmission.

\begin{figure}[tb]
\includegraphics[width=80mm]{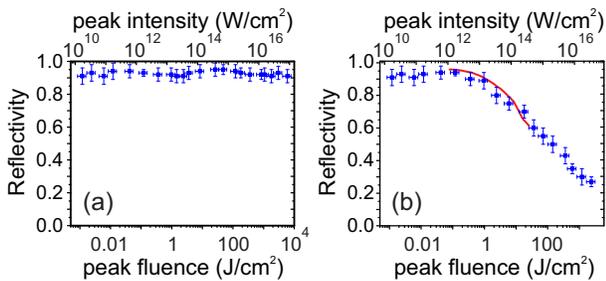}
\caption{ The measured dependence of the copper reflectivity on the incident peak intensity and fluence at (a) vacuum ($10^{-5}$\,Bar) and (b) air conditions. The solid curve in (b) shows the results of Kirkwood et al. \cite{PRB79-144120}. }\label{fig4}
\end{figure}

The peak intensity in the capillary experiment is limited by the available pump energy and does not reach the relativistic value required for the plasma wakefield acceleration. As there is a risk of transmission decrease due to the reduced metal reflectivity at high intensities \cite{PoP13-053114}, and experimental evidences of that in slightly different conditions are reported \cite{EPL55-334,LPB23-391,PRB79-144120}, the copper reflectivity at high intensities is directly measured using the same laser system. The experimental setup is shown in Fig.\,\ref{fig3}(b). The optical quality polished copper plates with nearly the same quality as the capillary walls are used in these experiments. The target is mounted on the commercial three-dimensional translation stage and irradiated at $45^\circ$ angle of incidence with s-polarized laser pulses in the single shot mode. After each shot, the target is moved in the direction parallel to the target surface, so that each laser pulse interacts with a fresh spot on the sample. The pulse energy is controlled by the pump energy, so no filters that could introduce nonlinear distortions are used. The copper sample is placed at the focus of the 38\,cm focal length lens with an accuracy of 1\,mm or better, which is within the Rayleigh range. The detectors are calibrated at low pulse energies (in the absence of ablation and surface damage) with a high-reflectivity dielectric mirror mounted instead of the copper target. To avoid systematic errors, several series of measurements in different parts of each of several samples are performed. The averaged results are shown in Fig.\,\ref{fig4}(a). No decrease of the reflectivity is observed up to the peak intensity of $10^{17}\,\text{W/cm}^2$.

The reflectivity weakly depends on the incidence angle for s-polarization \cite{PRL64-1250,EPL55-334}, thus the obtained experimental results are not in line with Refs.~\cite{EPL55-334,LPB23-391,PRB79-144120}. The discrepancy with Ref.~\cite{EPL55-334} may be due to different wavelength and material. The results of Ref.~\cite{PRB79-144120} are closely reproduced in air at 1\,Bar pressure [Fig.\,\ref{fig4}(b)], while in helium with the pressure up to 1\,Bar no decrease of the copper reflectivity is observed at the laser intensity of $10^{17}\,\text{W/cm}^2$. This suggests that the surrounding gas has a strong effect on the reflectivity, though the model used in Ref.~\cite{PRB79-144120} does not include the gas as a contributing factor. The difference with Ref.~\cite{LPB23-391} may be caused by the influence of the pulse contrast, which was rather low there ($10^{-5}$).

The intensity of $10^{17}\,\text{W/cm}^2$ on the capillary wall corresponds to about $10^{19}\,\text{W/cm}^2$ on the axis \cite{PoP13-053114}, so the experiments give promise that the transmission of relativistic intensity pulses through copper capillaries could be high, about 70\% for 20\,mm long capillaries, and the wakefield acceleration in plasma-walled capillaries is possible.

\begin{figure}[tb]
\includegraphics[width=80mm]{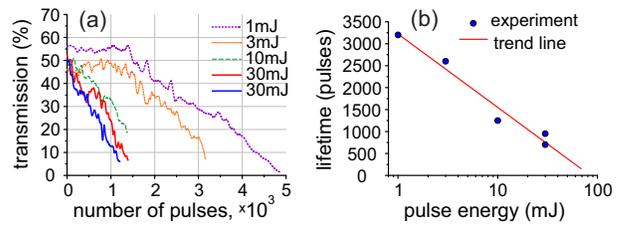}
\caption{ (a) Reduction of the absolute transmission with the number of transmitted pulses for different pulse energies. (b) Dependence of the capillary lifetime on the pulse energy. }\label{fig5}
\end{figure}

Another important question is related to capillary degradation and the capillary lifetime. To study this, the transmission is measured as a function of the number of transmitted pulses for various pulse energies [Fig.\,\ref{fig5}(a)]. The pulses follow at the repetition rate of 10\,Hz. The dependence of the capillary lifetime on the pulse energy is shown in Fig.\,\ref{fig5}(b). Here the lifetime is determined as  the number of pulses which causes the twofold reduction of the transmission. Extrapolating the data in Fig.\,\ref{fig5}(b) indicates that the lifetime of the copper capillary is about one pulse at pulse energies greater than 100\,mJ.

\begin{figure}[tb]
\includegraphics[width=80mm]{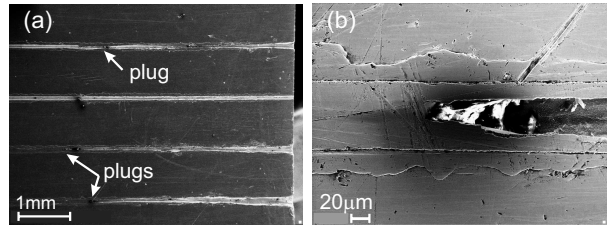}
\caption{ SEM images of the plugs in capillaries (a) and the enlarged origin of a plug (b). The plate with grooves is shown; the capillary entrance is on the right side of the photos. }\label{fig6}
\end{figure}

To determine the cause of the transmission decrease, the used capillaries are examined with the microscope. SEM images of the capillaries are shown in Fig.\,\ref{fig6}. Along with the expected degradation of the capillary entrance, plugs at the distance of 2--4\,mm from the entrance are observed. The plugs are about 1\,mm long and block the capillary completely. The plug, obviously, cannot block the very first pulse, so this mechanism of transmission degradation does not work for single pulses even at relativistic intensities that correspond to pulse energies greater than 0.5\,J. Consequently, the ablating metallic capillaries can find their use for laser wakefield acceleration only in a single shot regime, which means the capillaries must be easily produced and replaced for each laser shot.

To conclude, the critical elements of laser wakefield acceleration in plasma-walled capillaries are investigated up to the pulse intensities accessible with the used experimental setup. A technology of capillary production is developed which potentially enables mass replication of capillaries. These capillaries show the relative transmission as high as 70\% over the distance of about 15 Rayleigh lengths for the on-axis pulse intensities up to $10^{17}\,\text{W/cm}^2$. The special study of the copper reflectivity indicates that earlier observations of the reflectivity reduction at high pulse intensities may not be extrapolated to the laser parameters of interest which are characterized by a short pulse duration and a high contrast. The main cause of capillary degradation is found to be the plug formation. This effect may be material-dependent and needs further studies. In general, no stoppers are found on the path to guiding higher intensity pulses by plasma-walled capillaries for the laser wakefield acceleration.

This work is partly supported by the RAS Program of Basic Research ``Extreme Laser Radiation: Physics and Fundamental Applications''. The authors are grateful to Ya.\,L.\,Lukyanov for SEM image taking and to V.V.Ershov for capillary design and technology development.

\end{document}